\documentclass[12pt]{article}
\usepackage[margin=0.7in]{geometry}
\usepackage{float}
\usepackage{amsmath}
\usepackage{amssymb}
\usepackage{graphicx}
\usepackage{cite}
\usepackage{color}
\usepackage{dcolumn}
\usepackage{bm}
\usepackage{newfloat}
\DeclareFloatingEnvironment[name={SI Figure}]{suppfigure}
\DeclareFloatingEnvironment[name={SI equation}]{suppequation}
\DeclareFloatingEnvironment[name={SI Table}]{supptable}
\begin{document}

\begin{flushleft}
{\large
\textbf{Social patterns revealed through random matrix theory}
}
\\
Camellia Sarkar$^{1}$ \& Sarika Jalan$^{1,2,\ast}$
\\ 
\it ${^1}$ Centre for Biosciences and Biomedical Engineering, Indian Institute of Technology Indore, Simrol, Indore 452020, India\\
\it ${^2}$ Complex Systems Lab, Discipline of Physics, Indian Institute of Technology Indore, Simrol, Indore 452020, India\\

$\ast$ Corresponding author e-mail: sarika@iiti.ac.in
\end{flushleft}

\begin{abstract}
Despite the tremendous advancements in the field of network theory, very few studies
have taken weights in the interactions into consideration that emerge naturally in all real world
systems. Using random matrix analysis of a weighted social network, we demonstrate the profound impact of weights in interactions on emerging structural properties. The analysis reveals that randomness existing in particular time frame affects the decisions of individuals rendering
them more freedom of choice in situations of financial security. While the structural organization
of networks remain same throughout all datasets, random matrix theory provides insight into interaction pattern of individual of the society in situations of crisis. It has also been contemplated
that individual accountability in terms of weighted interactions remains as a key to success unless
segregation of tasks comes into play.
\end{abstract}

Apprehending various physical systems under the framework of networks has been in practice since a couple of decades 
\cite{Strogatz,Barabasi_2004}. Recent years have realized the need of this framework in understanding a myriad of social phenomena in various 
disciplines ranging from psychology to economics \cite{Borgatti_2009} for instance, in understanding spread of behaviour across a 
population \cite{Centola_Aral}, assessing organizational performance \cite{Thomas_2012}, outbreak detection of disease propagation \cite
{Eubank_2004}. In this Letter, we delve in to the intricacies of society using Bollywood, the largest film industry 
of the world \cite{Focus_2010} as a model. Being a mixture of art, business and entertainment, Bollywood, with 100 years 
of cinematic heritage has emerged as a key artifact providing a host of insights into shifting ideals, fantasies, lifestyles, 
opinions and views of the ever-evolving Indian society \cite{Bollywood_book}.  Making nearly one thousand feature films and fifteen hundred short films per year, the Indian film industry is the largest in the world. Cinemas in the UK have been screening Bollywood movies since the 
late 1960s and the spread of Indian population overseas renders Bollywood stars to gain 
popularity both home and abroad \cite{SM}. Bollywood provides a legitimate ground to capture the dynamically evolving
 nature of the diasporic society, where definitions of lead actors and their credentials of success was captured in terms of the payoff and overlap 
of the actors and the prevailing gender disparity in the society was emphasized upon \cite{SJ_Plosone_2014}.

Although Hollywood, a much smaller counterpart has been investigated using various statistical techniques \cite{Cattani}, analysis over an evolving time scale was not emphasized upon and impact of interaction strengths on social behaviour has also been largely ignored despite the realization of importance of weights in understanding system dynamics \cite{Barabasi_PRL_2001}.
For example, importance of weak ties has been emphasized in maintaining integrity of social systems \cite{Granovetter}. Further, the strength of interactions have been found to play a pivotal role in analysis 
of many other systems such as food-web structure, metabolic networks, 
scientific collaboration ties, air transportation networks, internet traffic \cite{Weighted}, suggesting that weights in interactions should be incorporated in order to have a more comprehensive picture of the structural organization of a system, hence forming a fair
 ground for us to investigate the weighted Bollywood networks.

This Letter, using random matrix theory (RMT), an involved mathematical tool which has 
demonstrated its success in unveiling crucial properties of various complex systems ranging from many body quantum systems \cite
{Weidenmuller_2007} to biological ones \cite{Potestio_PRL_2009}, analyses the weighted social networks and reveals that weights play a 
decisive role in extracting crucial aspects of human behaviour. Furthermore, we demonstrate that structural analysis of weighted networks unfolds individual credentials to impact success much more than the influence of the fraternity they work in. 
Interesting revelations as well as differences from the unweighted networks appear from the spectral analyses of the weighted Bollywood networks using RMT, which provide an insight in to the preferences of the society based on randomness.

\section*{Methods}
\subsection*{Construction of weighted Bollywood co-actor networks} 
We divide the massive Bollywood data, collected for a span of 60 years from the movie repository websites 
\cite{website}, into five-year window periods. Time intervals on one hand should be large enough to capture specific properties
of the society as well as to yield statistically significant results and on other hand should be reasonably
small to capture the changes. Since the model system is based on rapidly changing society \cite{Bollywood_book}, we find intervals of five years to be an apt time frame. The networks for each of the datasets are constructed where actors are the nodes and connections with their co-actors are the links \cite{SJ_Plosone_2014}, an additional attribute considered here being weights. By weight we refer to the number of movies an actor has co-acted with another actor in a particular five-year time span. The adjacency matrix, A of the networks thus generated are given as:
\begin{equation}
A_{\mathrm {ij}} = \begin{cases} w_{\mathrm {ij}}~~\mbox{if } i \sim j \\
0 ~~ \mbox{otherwise} \end{cases}
\label{adj_wei}
\end{equation}
where $w_{\mathrm{ij}}$ is the number of times (movies) actors i and j act together in a particular span.
Properties of networks constructed with different datasets are provided in Table~\ref{table}.

\subsection*{Structural parameters} 
Clustering coefficient (CC) of a node is defined as the ratio of the number of links between the neighbours of the node to the possible number of links that could exist between the neighbours \cite{Newman_2003}.
Betweenness centrality ($\beta_c$) for a node {\it i} is defined as \cite{Newman_2003}
\begin{equation}
\beta_{c} = \sum_{St} \frac{n^i_{St}}{g_{St}}
\label{betweenness}
\end{equation} 
where $n^i_{St}$ is the number of shortest paths from $s$ to $t$ that passes through 
$i$ and $g_{St}$ is the total number of shortest paths from $s$ to $t$.

\subsection*{Spectral properties} 
The spectra of the corresponding adjacency matrix
is denoted by $\lambda_i = 1,\hdots ,N$ and 
$\lambda_1>\lambda_2> \lambda_3> \hdots > \lambda_N$. In RMT, it is customary to unfold the data 
by a transformation $\bar{\lambda_i}= \bar{N}(\lambda_i)$, where $\bar{N}$ 
is average integrated eigenvalue density \cite{Mehta_book}. In absence of any analytical form for $\bar{N}$, we perform unfolding by numerical polynomial fitting using the smooth part of the spectra obtained by discarding eigenvalues towards both the ends as well as degenerate eigenvalues, rendering the dimension of the unfolded spectra $N_{eff}$.
Using the unfolded spectra, we calculate spacings $s^{(i)}= \bar{\lambda}_{i+1} - \bar{\lambda_{i}}$ distribution $\rho(s)$ and fit it by the Brody distribution (Eq.~\ref{eq_brody}) characterized by the parameter $\beta$ \cite{Mehta_book} as follows:
\begin{equation}
P_{\beta}(s)=As^\beta\exp\left(-\alpha s^{\beta+1}\right)
\label{eq_brody}
\end{equation}
where $A$ and $\alpha$ are determined by the parameter $\beta$ as: 
$A =(1+\beta)\alpha, \, \, \, \alpha=\left[{\Gamma{\left(\frac{\beta+2}{\beta+1} \right)  }}\right]^{\beta+1}$.

We analyze the long range correlations in eigenvalues using $\Delta_3(L)$ statistics
which measures the least-square deviation of the spectral 
staircase function representing average integrated eigenvalue density 
${N}(\bar\lambda)$ from the best fitted straight line for a finite interval of 
length $L$ of the spectrum \cite{Mehta_book} and is given by
\begin{equation}
\label{eq_delta3}
\Delta_3 (L;x) = \frac{1}{L} min_{a,b}   \int_x^{x+L} [N(\overline{\lambda})-a\overline{\lambda}-b]^2 d\overline{\lambda}
\end{equation}
where $a$ and $b$ are regression coefficients obtained after least square fit. Average over several choices
 of x gives the spectral rigidity, the $\Delta_3(L)$.
In case of GOE statistics, $\Delta_3(L)$ statistics depends logarithmically on $L$ given as:
\begin{equation}
\Delta_3(L)  \sim \frac{1}{\pi^2} \ln L
\label{eq_delta3_goe}
\end{equation}
As the value of $L$ increases, the number of data points for the same size of the network becomes less and we choose the value $L$ 
around $1/4$ of the unfolded spectra in order to perform statistical calculations.

\subsection*{Weights and success correlation}
We pick the top ten actors from the list of Filmfare award nominees \cite{SM}
for each span and compare their positions in degree order in unweighted networks to those of weighted networks. Further we calculate the ratio of their award nominations across consecutive spans. Occurrence of both these ratios greater than one simultaneously for consecutive spans for any actor  indicates that working more frequently with a confined set of actors as compared to the previous span is implicative in his (her) success. The value of the first ratio above one and that of the second ratio less than one implies that working with the same set of actors more often as compared to the preceding span does not assure success of an actor. 

\section*{Results and Discussions}
\subsection*{Hierarchical nature of weighted Bollywood networks} 
The size of the networks increase with time (Table~\ref{table}) indicating the growing success
of the Bollywood. The drastic change in the size during 1998-02 span combined with
the `industry' status conferred to the Bollywood \cite{Ray_2012} indicates that financial security
has an upper edge over artistic excellence. Almost constant value of $\langle k \rangle$ across 
60 years indicates the characteristic nature of the model system in turn reflecting
the overall connectivity of actors themselves.
Further, we find that the degree distribution exhibit power law for the weighted Bollywood networks as also observed for the
unweighted networks \cite{SJ_Plosone_2014} suggesting that overall contribution of weights does not bring about any significant change in the distribution of number of co-actors in the networks. 
\begin{table}[t]
\begin{center}
\caption{
Properties of weighted Bollywood networks of each 5 years block datasets. $N_{eff}$ is the effective dimension of the network. The \%$L_{0(unw)}$ and \% $L_{0(w)}$ represent the extent of $L_{0}$ up to which spectra follow GOE statistics, expressed in percentage terms for unweighted \cite{SJ_Plosone_2014} and weighted Bollywood networks respectively. `-' denotes that the $\Delta_3(L)$ statistics does not follow GOE statistics.}
\begin{tabular}{|c|c|c|c|c|c|c|}    \hline
{\small Time span} & {\small $N$} & {\small $\langle k \rangle$} & {\small $L_{0}$} & {\small $N_{eff}$}  & {\small \%$L_{0(w)}$} & {\small \%$L_{0(unw)}$} \\ \hline
53-57		& 788 	& 26	&	25	&	250	&	10	&	-	\\ \hline
58-62		&  827 	& 30	&    	16	&	250	&	6.40	&	-	\\ \hline
63-67		&  772 	& 35	&	10	&	280	&	3.57	&	6.16	\\ \hline
68-72		&  1036 & 47	&	11	&	350	&	3.14	&	-	\\ \hline
73-77		&  990	& 48	&	-	&	300	&	-	&	3.65	\\ \hline
78-82		&   968	& 45	&	12	&	325	&	3.69	&	4.32	\\ \hline
83-87		&  1335	& 45	&	46	&	440	&	-	&	3.95	\\ \hline
88-92		& 1465	& 45	&	15	&	450	&	3.33	&	4.39	\\ \hline
93-97		& 1314	& 42	&	28	&	200	&	14	&	2.38	\\ \hline
98-02		& 1878	& 46	&	44	&	250	&	17.60	&	2.04	\\ \hline
03-07		& 2935	& 37	&	18	&	300	&	6	&	1.74	\\ \hline
08-12		& 3611	& 30	&	43	&	300	&	14.33	&	1.46	\\ \hline
\end{tabular}
\label{table}
\end{center}
\end{table}
\begin{figure}
\centering{\includegraphics[width=0.6\columnwidth]{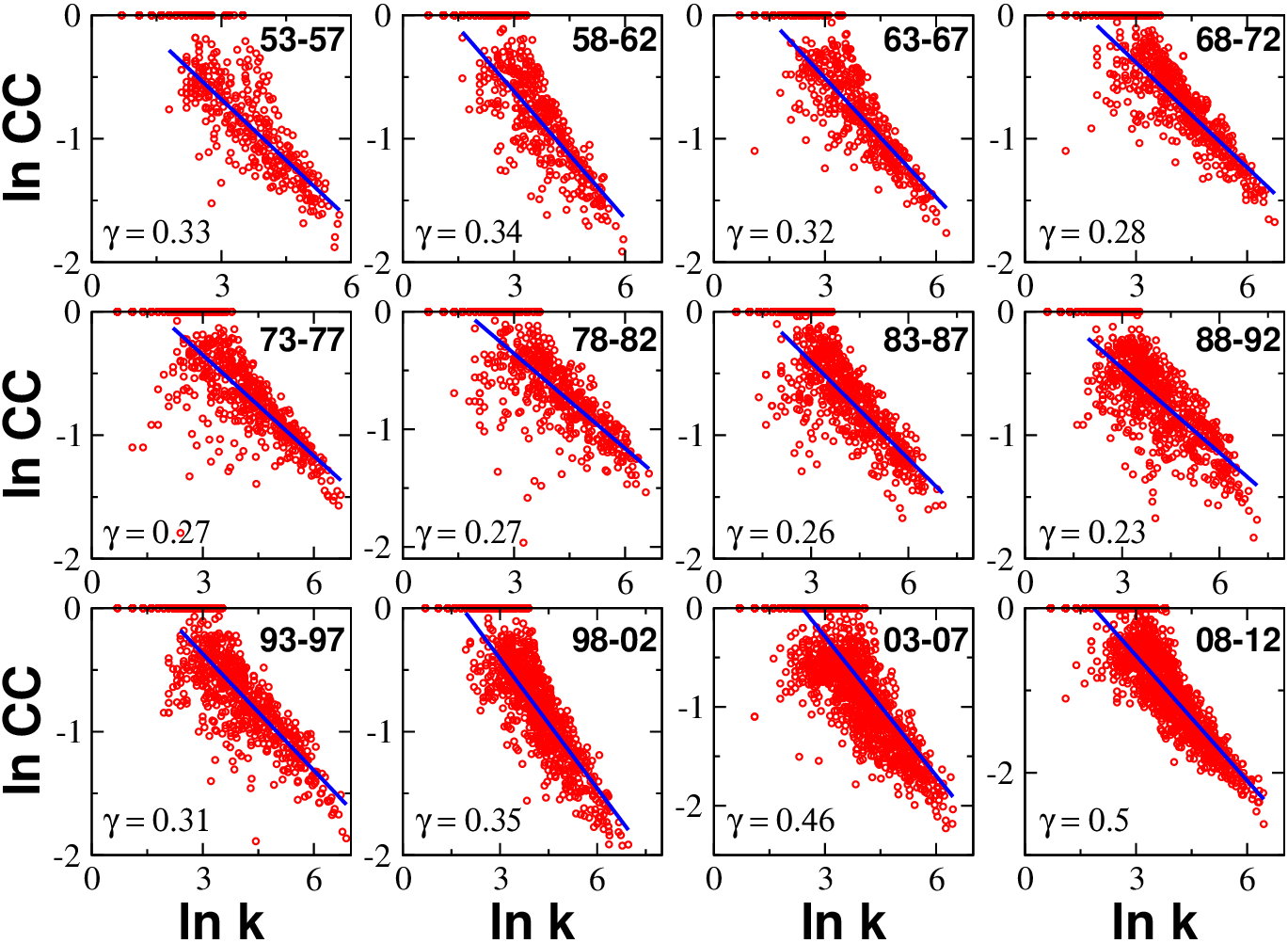}}
\caption{(Color online) Plots of normalized weighted degree and clustering coefficient (CC) over 1953-2012 time spans. The red circles represent the data points and the blue line marks their power law fit. `$\gamma$' refers to the exponent of power law.}
\label{lnk_lnCC}
\end{figure}
Next, we find negative correlation between degree-CC for all the datasets 
(Fig.~\ref{lnk_lnCC}), as also observed for other social networks \cite{Goh_2003}.
What follows from the method of network construction is that the new comers in the industry, who have acted in just one movie in a particular span will have their $CC =1$ at the time of their entry into the network (Bollywood). Further, though there are various algorithms \cite{Charo_Goh_Judd} known to yield networks with the power law degree distribution, the
preferential attachment here seems to be most logical one \cite{SJ_Plosone_2014}, leading low degree nodes to get associated with high degree nodes. A low degree node, even when it forms a subset of the large number of co-actors of a high degree node it is connected with, while entering into the system, cannot link with all the co-actors of the high degree node hence leading to lowering the CC of the high degree node and it can be comprehended that more the number of associated low degree nodes more will be the lowering in the CC of the high degree node. This depicts that in a society being associated with a large number of people and more number of less associated people 
(belonging to tightly knit groups \cite{Opsahl_2009}) might lead to poor clustering of oneself. 
Note that the power law degree-CC correlation (Fig.~\ref{lnk_lnCC}) might be arising due to the hierarchical nature \cite{Barabasi_2004}
of the Bollywood owing to several reasons such as the actors grouping together for low budget movies, directors preferences \cite{Bollywood_book} etc. The small value of $\gamma$ indicates less prominent hierarchical structure. Moreover, similar nature of degree-CC correlation across the datasets reflects the stationary nature of the Bollywood networks.

\subsection*{Strengths of Bollywood ties affect randomness} 
\begin{figure}
\centering{\includegraphics[width=0.6\columnwidth]{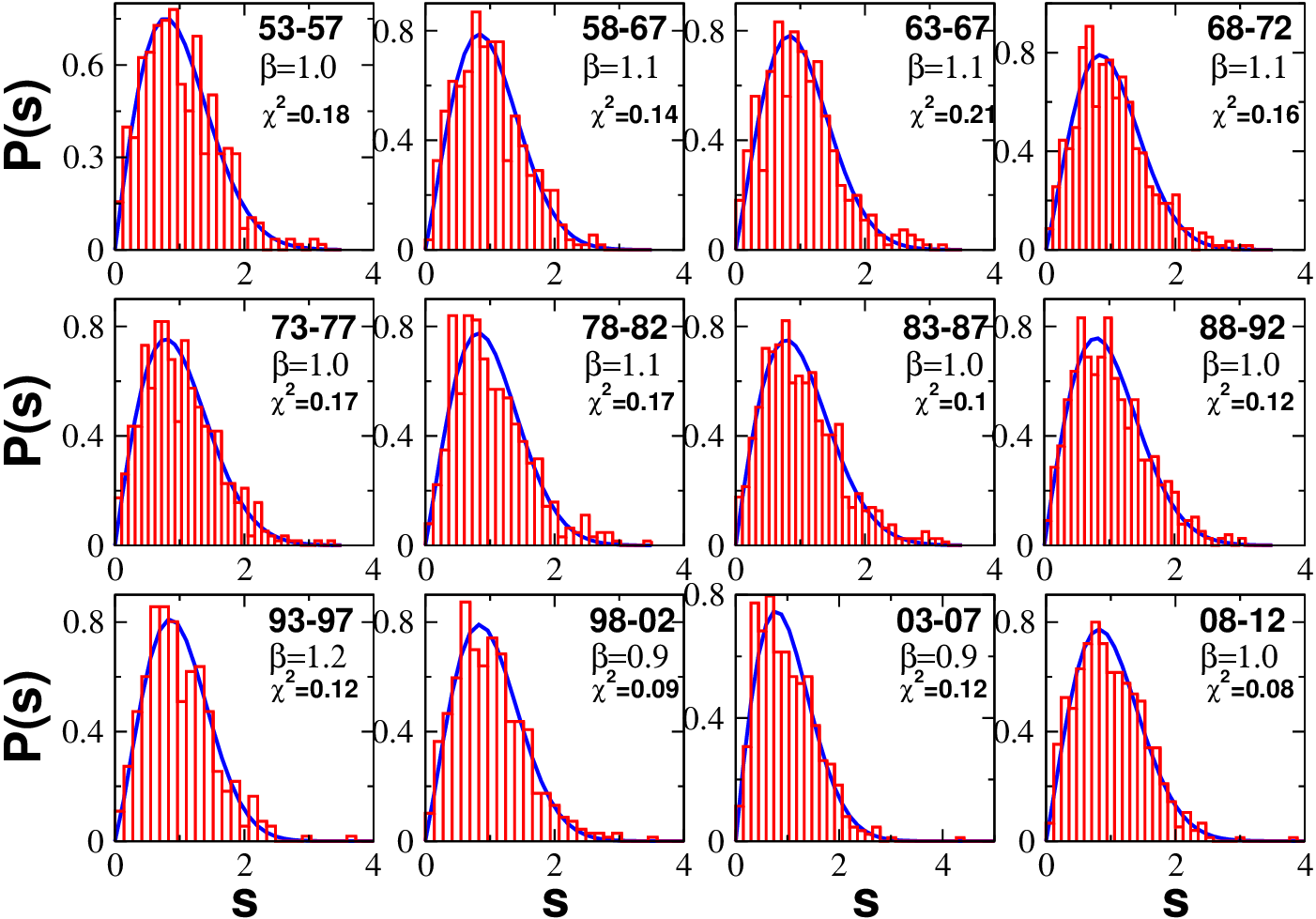}}
\caption{(Color online) Nearest-neighbour spacing distribution $P(s)$ of the adjacency matrix of weighted Bollywood
networks for 1953-2012 datasets. Histograms are numerical results and solid lines represent the NNSD of GOE. The $\chi^{2}$ values provide a measure of the error between hypothesized model and observed data. Lower values of $\chi^{2}$ values (close to 0) show high degree of goodness of fit.}
\label{NNSD}
\end{figure}

The model considered here being based on a rapidly changing society \cite{Bollywood_book}, provides an apt platform to understand the impact of this change on human behaviour across time. The structural analyses, while demonstrating important universal properties, fail to discern time varying transitions, leading us to go beyond structural analyses and we turn up with spectral analysis under RMT framework.
The nearest neighbour spacing distribution (NNSD) is one of the most popular technique in 
RMT which provides the information about short range correlations in eigenvalues and
form a basis to understand universality in the corresponding spectra.
We find that the NNSD of the weighted Bollywood networks, when fitted with Eq.~\ref{eq_brody} yields value of $\beta$ close to one indicating that the spectra follow the universal GOE statistics of RMT (Fig.~\ref{NNSD}). This is not very surprising as NNSD for the unweighted Bollywood networks have also been observed to follow
GOE statistics \cite{SJ_Plosone_2014}. This universal GOE behaviour suggests that there exists some {\it minimal amount of randomness} in the weighted Bollywood networks, sufficient enough to introduce short range correlations in the eigenvalues, though it 
does not quantify the amount of randomness existing. Importance of randomness in the establishment and the 
conservation of complexity in social structures has been investigated deploying interaction dynamics of a population of wild house mice \cite
{random_mice}. Both unweighted and weighted Bollywood networks following this universal behaviour, is indicative of the notion that 
irrespective of the strengths in Bollywood ties, their underlying networks possess some randomness which might be instrumental in 
conferring robustness \cite{SJ_Plosone_2014} to the system. 

\begin{figure}
\centering{\includegraphics[width=0.6\columnwidth]{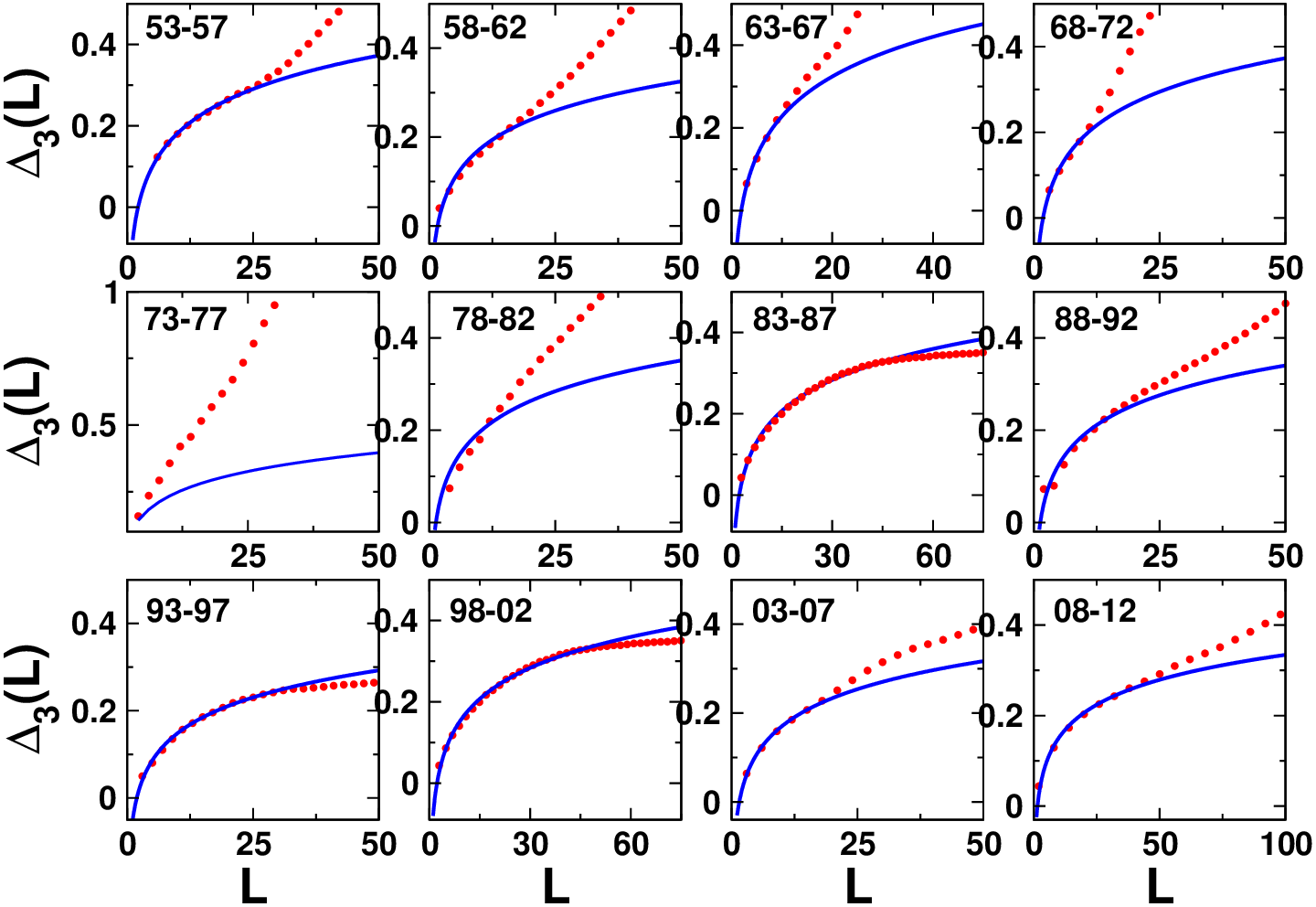}}
\caption{(Color online) $\Delta_3(L)$ statistics of the weighted Bollywood networks for 1953-2012 datasets. We plot the values of $L$ of $\Delta_3(L)$ statistic for the data together with that of the GOE. The extent up to which our data points are 
concurrent with the GOE fit (the red circles follow the blue line) provides the value of $L_0$.}
\label{delta3}
\end{figure}

The NNSD only captures short range correlations in the eigenvalues. In order to gain a deeper insight to the impact of weights in randomness we analyze long range correlations in the eigenvalues using the $\Delta_3(L)$ statistics (Eq.~\ref{eq_delta3}).
We find that the weighted Bollywood networks for different spans follow random matrix prediction of GOE statistics (Eq.~\ref{eq_delta3_goe}) up to a certain value $L_{0}$ and deviates afterwards (Table~\ref{table} and Fig.~\ref{delta3}). The value of $L_{0}$, for which the statistics follows universality,
has been used as measure of randomness in networks \cite{SJ_EPL_2009}, further aiding us to deduce properties
of the model when weights are considered in defining the network.
 What follows that the datasets 1973-77 and 1983-87 failed to follow random matrix predictions while 1998-02 dataset possessed the maximum amount of randomness among all other datasets in weighted Bollywood networks. 

In the following, we demonstrate how inclusion of weights in the networks not only changes the statistical properties yielded through long range correlations, but also provides an insight in to the hidden patterns in the underlying system. Table~\ref{table} shows that the length $L_{0}$ for which individual sets follow RMT (Eq.~\ref{eq_delta3_goe}), is remarkably different for the weighted ($\%L_{0(w)}$) and unweighted ($\%L_{0(unw)}$) Bollywood networks (Table~\ref{table}). Except for 1963-67 and the spans from 1973-1992, all other datasets witness increase in randomness of Bollywood networks on considering weights. Randomness in unweighted networks solely relied on the distribution of the zero and one elements and the statistics they yield. Leaving the zero elements in the matrices unaltered, the one elements are replaced with their corresponding weights in case of weighted networks, thus introducing an extra dimension to calculation of spectral properties of the networks and could be one of the reasons which contribute to increase in randomness. The five datasets defying this argument depict a completely different picture indicating that working in closed circles might have rendered the underlying networks to adopt some structure. We predict that some kind of defined priorities might have existed in those time frames in Bollywood. We found that in 1973-77 span we witness director Hrishikesh Mukherjee to have preferred the popular duo Amitabh Bachchan and Jaya Bhaduri in three out of six movies that he had directed in that span. Similarly, director Yash Chopra chose the duo Amitabh Bachchan and Neetu Singh twice out of the four movies that he had directed in that span \cite{SM}. In the spans where decrease in randomness was witnessed, wars, communal riots and resulting conflict of Bollywood stars with state government \cite{SM} were observed which led to financial crisis \cite{Brenda}, leading the top directors of those spans to give preference to certain sets of actors over others. 

Note that while networks with different size might have same amount of randomness (measured through $\Delta_3$ statistic) \cite{SJ_PRE_2007}, varying amounts of randomness might be observed in networks of same size and average degree \cite{SJ_EPL_2009}. Hence, the variations in the $\Delta_3(L)$ statistic for different datasets are not due to the variations in size, and they capture changes in interaction patterns due to the inclusion of weights in the underlying networks.

\subsection*{How centrality affects proximity of co-actors} 
\begin{figure}
\centering{\includegraphics[width=0.6\columnwidth]{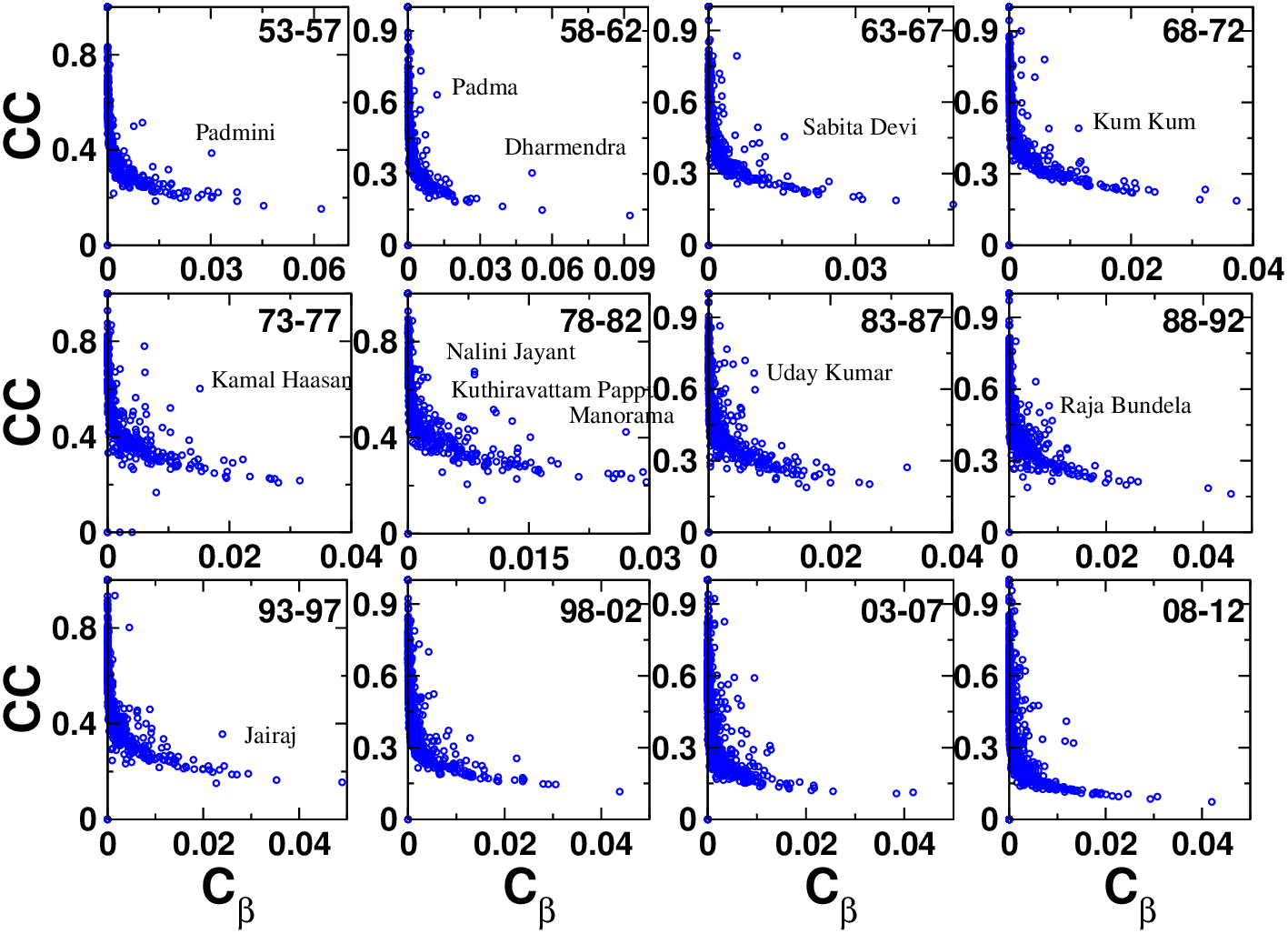}}
\caption{(Color online) Plots of normalized clustering coefficient (CC) against normalized betweenness centrality ($\beta_c$) over 
1953-2012 time spans. The blue circles represent the data points of the plot.}
\label{CC_BC}
\end{figure}

Through above analysis we observe that spectra has proved its credibility in discerning universal social behaviour and capturing crucial properties of human behaviour like selective preference to known candidates under crisis. We probe further to investigate the impact of the changing society on human behaviour relating the spectra with the time-bound events happening in our model system using betweenness centrality ($\beta_c$) as a tool.
In our analysis of weighted Bollywood networks, $\beta_c$ presents a negative correlation with CC, although few nodes appear having reasonably high $\beta_c$ yet high CC (Fig.~\ref{CC_BC}). These nodes, apart from connecting different 
Bollywood circles might be instrumental in connecting their co-actors within their domain. Among them, few of the actors like Kamal 
Haasan, Padmini, Manorama, Uday Kumar, Jairaj are established actors of other regional film industries. Owing to their successful 
realm in their respective regional film industries, they might be playing a central role in connecting their Bollywood co-actors. Combining 
with the functional importance of these nodes and their long span in the industry \cite{SM}, our analysis suggests that they are not arising due to random fluctuations. Corresponding configuration model exhibiting a smooth $\beta_{c}$-CC anti-correlation with no scattered data points, strengthens our argument. No such nodes have appeared in the datasets 1998-02 
onwards (Fig.~\ref{CC_BC}) indicating that the corresponding networks do not possess central nodes which 
are clustered. Incidentally 1998-02 dataset has been revealed in $\Delta_3(L)$ statistics analyses of weighted Bollywood networks to be 
the most random based on the value of $L_{0}$ (Table~\ref{table} and Fig.~\ref{delta3}). These observations direct us to suggest that the Bollywood 
actors, on account of experiencing more financial security owing to endowment of `industry status' to Bollywood in 1998 \cite{Ray_2012}, 
tend to enjoy the freedom of working randomly with the co-actors of their choice devoid of any kind of bias emphasizing that financial 
security is a key feature driving different strata of society \cite{Bollywood_book}.

\subsection*{Impact of weights on success of Bollywood superstars} 
\begin{figure}
\centering{\includegraphics[width=0.6\columnwidth]{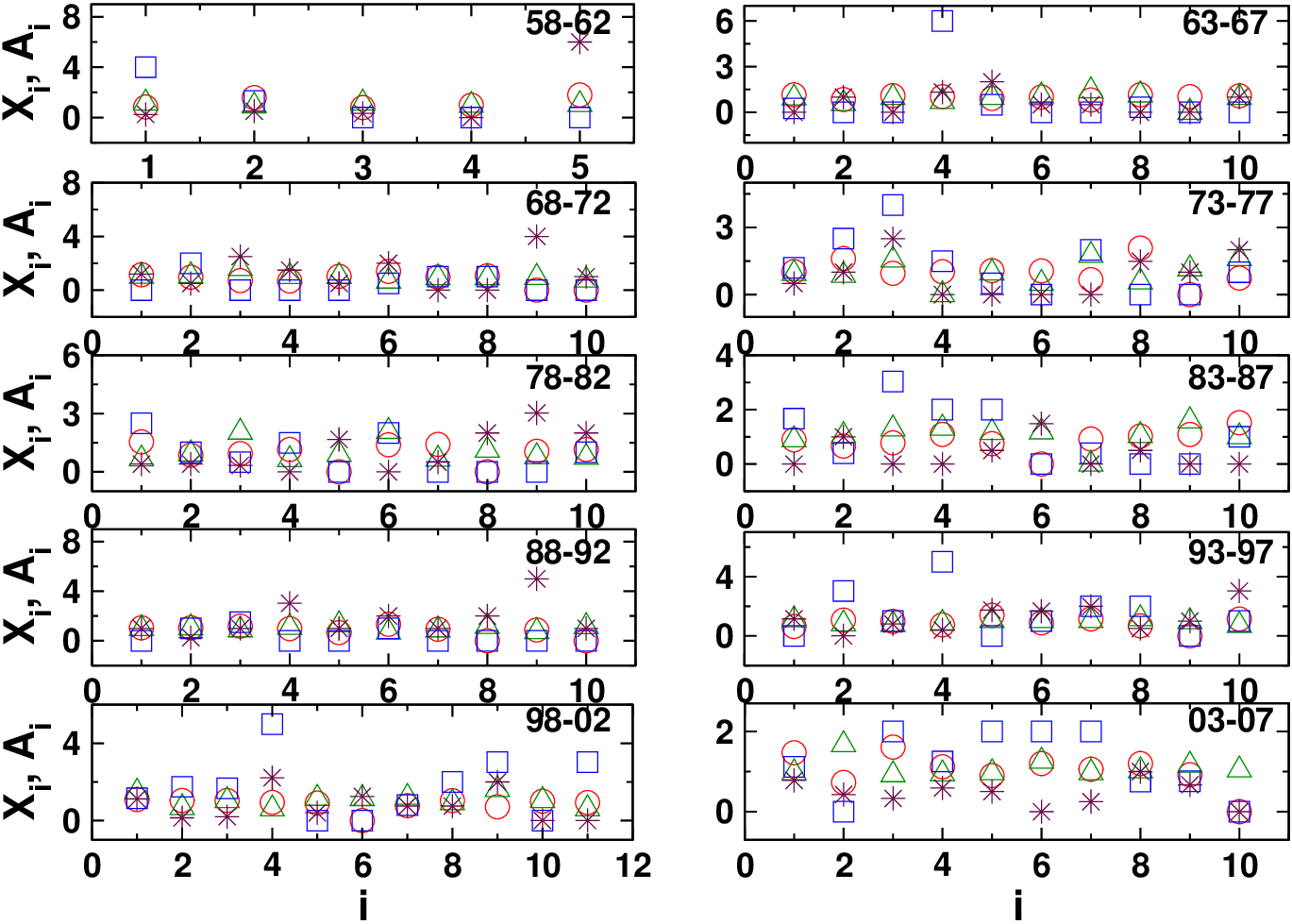}}
\caption{(Color online) X and A of the ten Bollywood actors (i) having maximum Filmfare award nominations for present versus preceding span (circles and squares) and succeeding versus present span (triangles and stars), representing X and A respectively in each case. Here `X' refers to the ratio of the degree order in unweighted and weighted Bollywood networks across consecutive time spans. `A' refers to the ratio of the award nominations across consecutive spans.}
\label{X_A}
\end{figure}

Apart from holding a rich database which has proved its efficiency in understanding the social behaviour of the time varying society, our model also provides an additional measure which facilitates the assessment of success using award nominations for which we create a database of Filmfare nominees for 60 years segregated in to five-year time spans \cite{SM}. A popular notion evident in organizations demonstrates that group facilitation encourages coordination and leads to successful ventures \cite{Team_success}. Bollywood, acting as a mirror of the diasporic society \cite{Bollywood_book} provides us an opportunity to investigate whether working in closed circles (drawn in the form of weights) is implicative in success of individual actors or is it driven only by calibre and brilliance of actors. This is expected to draw a finer picture of the dynamically evolving society. 
We conduct a thorough analysis of the change in positions in degree sequence of the Bollywood stars for the weighted networks as 
compared to the unweighted ones. We find that in none of the datasets, either of these trends was consistently portrayed by majority of the 
leading successful actors of the respective era (Fig.~\ref{X_A}). This observation is quite counter-intuitive to the aforementioned popular 
notion. This inconsistency might be due to the fact that an 
actor's success does not solely depend on the credentials of his (her) co-actors but on his (her) own artistic excellence. For instance, the 
movie ``Mangal Pandey: The Rising'' was declared a below average movie by the Box Office India but its lead actor Aamir Khan 
gained critical acclaim leading him to win the Filmfare Award for Best Actor in 2006 \cite{SM}. A very common example of group activity 
{\it i.e.} cooperative learning methods, where every member of the group is assigned a sub-task, stands successful, only if group rewards 
are provided \cite{coop_learning}. Since in Bollywood, there is no division of task and every actor has the whole sole responsibility of 
making the movie hit, group accomplishment does not stand valid. Keeping in view the amount of popularity Bollywood celebrities have 
gained over the years both home and abroad and the sustenance of Bollywood in adverse situations, one can safely assume that Bollywood 
acts as a representative unit of the society. This leads us to propose that in a social system, unless sub-tasks are assigned and group 
recognition is bestowed upon, individual accountability remains as the key to success.

\section*{Conclusion}
To summarize, the universality in structural properties such as degree distribution and degree-CC  correlations
indicate time independent or stationary nature of the Bollywood.
Further, degree distribution following power law
has been known to be important for robustness of underlying system \cite{Barabasi_2004}, which in present context reflects the sustenance of our model system across time 
in situations of dire socio-economic crisis \cite{report}. Changes in the society 1998 onwards as reflected in $\beta_c$-CC correlation plots 
are related with the endowment of `industry' status to Bollywood, whereas changes in the society during 1963-67 and 1973-92 as reflected in $
\Delta_3(L)$ statistic, relates consecutive wars, communal riots and conflict of Bollywood actors with the state Government.

Using RMT, we demonstrate that
weights in interactions have a profound impact on the social structure. Although weights emerge naturally in real world systems, they have 
less often been investigated in real datasets despite significant advancements in network theory in the past two decades. In our analysis 
randomness in the weighted social network in certain time spans which witnessed social and financial chaos, sheds light on how the society 
is structured relating it with evidences where certain actor pairs were supported over others by the leading directors of the respective era. 
Although this conveys that the weighted model networks under consideration depict some social order, universality in NNSD is indicative of 
the {\it minimal amount of randomness}, implicated to be important in conferring robustness. Inconsistency in the statistics of success of 
individual node of the network correlated with their performance in defined circles based on weighted interactions has led us to suppose 
that self excellence appears more dominant than group coherence. Further, spectral and structural analysis based on proximity and 
centrality reveals importance of financial security driving the society. Although the model considered here is known to predominantly 
capture the behavioural aspects of the social framework of a part of the global population \cite{global}, the view that despite inherent 
population heterogeneity, human mobility portrays a deep-rooted regularity \cite{Barabasi_Nature_2008}, leads us to believe that the 
social behaviour revealed through our analysis would stand universally predictive. Furthermore, importance of weights in prediction of social 
patterns being revealed in the light of RMT, adds one more milestone to the success list of RMT.

\section*{Acknowledgements}
This work was funded by the DST grant SR/FTP/PS-067/2011 and CSIR grant 25(0205)/12/EMR-II. SJ thanks Sanjay Kumar for discussions on the model system and Stefano Boccaletti for encouraging discussions on applicability of RMT during CDSA, Kolkata.
\pagebreak

\newpage

\cleardoublepage
{\LARGE \bf Supplementary Material}
\vspace{1cm}
\section*{Bollywood}
Bollywood is the largest film industry in the world which accounts for around $40\%$ of the total revenues of the Indian film industry according to the CII-KPMG report of 2005 \cite{CII_KPMG}. A total of 8931 movies have been documented so far in Bollywood from 1913 to
2012. Bollywood is not only confined to the Asian sub-continent but has 
global outreach [Ref. 8 of manuscript]. The large fraction of Indian emigrants \cite{Overseas} invites the screening of Bollywood movies across the globe in various countries. For example, in countries like UK, Morocco and many more, the Bollywood actors see huge enthusiasm from the people over there \cite{Bol_baatein,Bol_Morocco}. 
\section*{Method for construction of networks}
From the Bollywood data collected for a period of 60 years spanning from
1953 to 2012 from the movie repository website {\it www.bollywoodhungama.com}, names of all the movies and their corresponding information are extracted using Python code. Initially we document names of all the films as per their chronological sequence (latest to
oldest) from the websites by incorporating the desired URL \cite{URL} in the code along with a built-in string
function which takes the page numbers (932 pages in “Released before 2012” category and 24 pages in
“Released in 2012” category) as input. Each film of every page bears a unique cast ID in the website,
navigating to which via “Movie Info” provides us with complete information about the film. In the Python
code, we store the unique cast IDs of films in a temporary variable and retrieve relevant information using
appropriate keywords from respective html pages. We also manually browse through the websites {\it www.imdb.com} and {\it www.fridayrelease.com} in order to collect any yearwise missing data, if any. Henceforth we merge the data from all the websites
and omit repetitions. Harvesting the complete data includes manual verification, formatting, removal of typos and compilation of data. Considering the
rapidly changing nature of the Bollywood network, we segregate the curated massive Bollywood data in to
datasets each containing movie data for five-year window periods.

We create the database of all actors and assign a unique ID number
to each actor in every span which we preserve throughout our analysis. We get rid of ambiguities
in spellings of names of actors presented in different websites by extensive thorough manual search and
cross-checking to avoid overlapping of information and duplication of node identities while constructing
networks. Seeing a $9\%$ growth from 2009 to 2010 according to the FCCI-KPMG report of 2011 \cite{FICCI} and a further $11.5\%$ growth in 2011 in comparison to 2010 according to the  2012 report \cite{FICCI}, it is a sphere which sees blazingly fast growth, leading us to expect drastic changes in small time frames. Tracking by their unique ID numbers, we create a co-actor database for each
span where every pair of actors who had co-acted in a movie within those five years are enlisted.
\section*{Degree distribution of weighted Bollywood networks}
The degree distribution of the weighted Bollywood networks are observed to follow power law (Fig.~\ref{deg_dist}). For some of the datasets, power law exponent ($\eta$) comes out to be less than two which might be due to the finite size effect.
\begin{suppfigure}[ht]
\centering
\includegraphics[width=0.6\columnwidth]{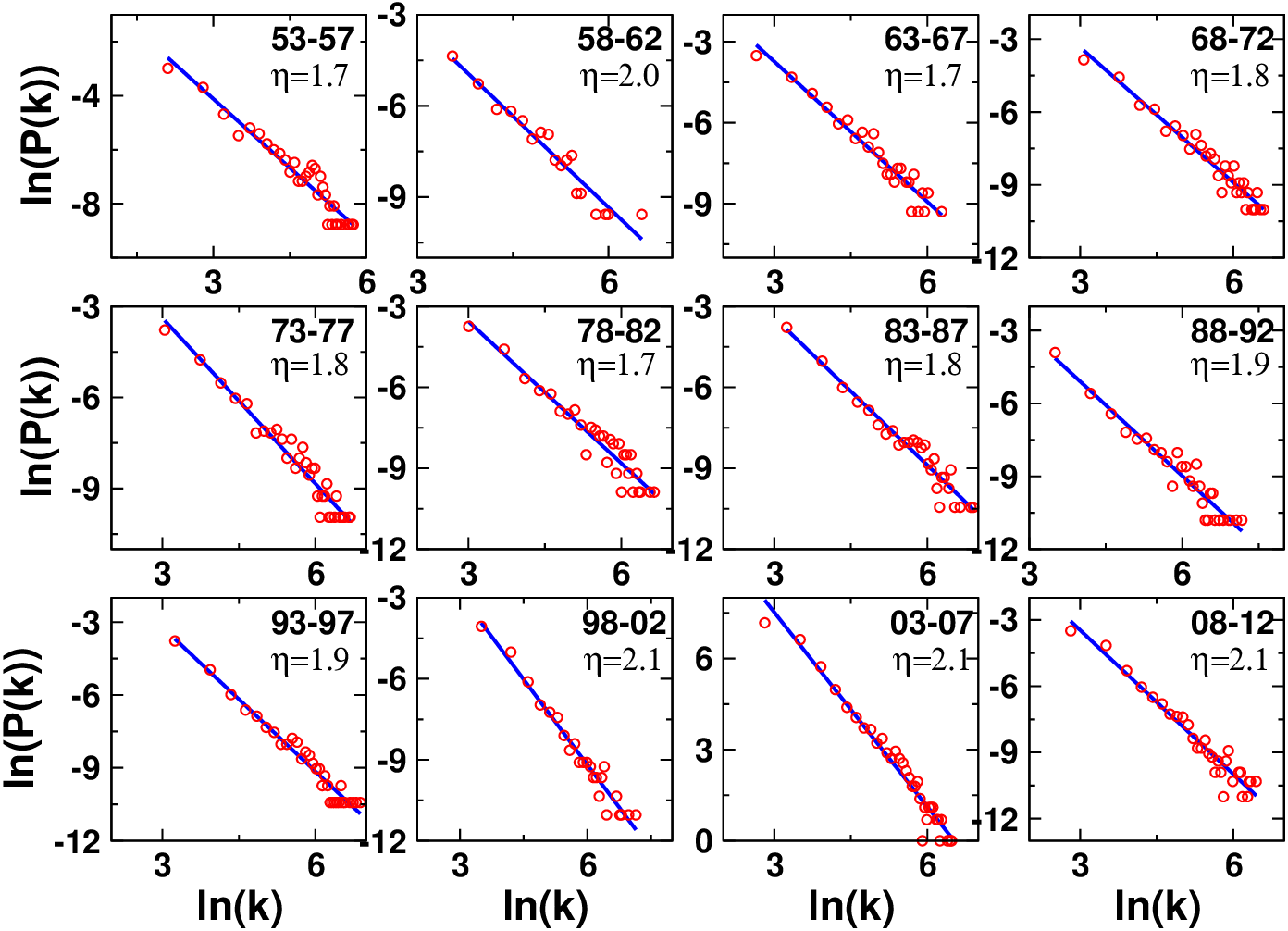}
\caption{(Color online) Degree distribution of weighted Bollywood networks for 1953-2012 time spans.}
\label{deg_dist}
\end{suppfigure}
\section*{Method of unfolding}
Unfolding is usually done by a transformation $\bar{\lambda_i}=  \bar{N}(\lambda_i)$, 
where $\bar{N}(\lambda) = \int_{\lambda_{min}}^{\lambda} \rho(\lambda') d\lambda'$
is the average integrated eigenvalue density [Ref. 19 of manuscript].
\section*{Statistical significance of the plots}
The statistical significance of the plots are presented in the form of $\chi^{2}$, which measures the extent of deviation of the observed data from the hypothesized model \cite{Gumbel}.
\section*{Configuration model}
We construct a configuration model \cite{Molloy} with the same degree sequence as of a real dataset and plot the $\beta_{c}$-CC correlation for ten different random realizations representing fluctuations (Fig.~\ref{CC_BC_config}).  
\begin{suppfigure}[ht]
\centering
\includegraphics[width=0.3\columnwidth]{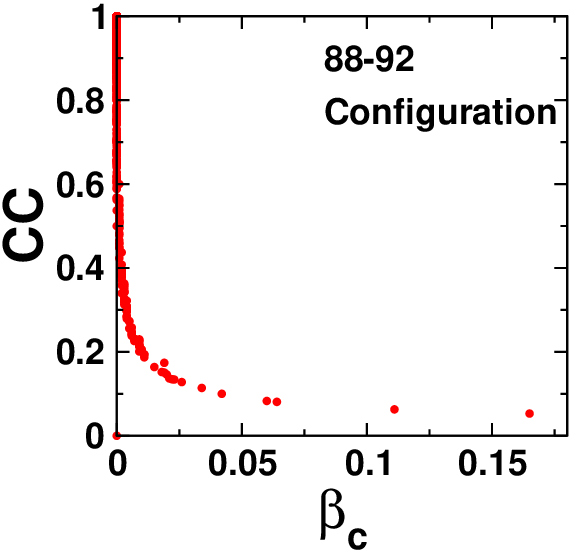}
\caption{(Color online) Plot of $\beta_c$ versus CC for a configuration model generated for a real data having the same degree sequence as the real data.}
\label{CC_BC_config}
\end{suppfigure}
\section*{Construction of Filmfare awards database}
Filmfare awards were first introduced by The Times Group \cite{Filmfare} after the Central Board of Film Certification (CBFC) was founded by Indian central government in 1952 to secure the identity of Indian culture. The reason we chose Filmfare Awards amongst all other awards is that it is voted both by the public and a committee of experts, thus gaining more acceptance over the years. Instead of the awards bagged we take into account the award nominations in order to avoid the interplay of some kind of bias affecting the decision of the CBFC committee in selecting the winner. By manual navigation through every year of Filmfare awards available on the web, we create database of all categories of Filmfare awards and extract their respective nominees chronologically from the html pages using Python codes. Henceforth, we use C++ codes to count number of times every actor is nominated in each five-year span. Thus we obtain a complete list of all actors in each span along with their number of Filmfare award nominations. Filmfare awards being rewarded 1954 onwards adds to the reason why we restrict our analysis pertaining to success of the Bollywood actors 1953-57 time span onwards.
\section*{Data pertaining to directors and actors of movies}
Table S1 accounts for the directors who were seen to prefer a set of actors over the rest in particular time spans (the names of directors of the relevant movies are retrieved from {\it www.imdb.com}). The corresponding movies and their actors have also been enlisted. 
\begin{supptable}[ht]
\begin{center}
\caption{List of directors who have preferred set of actors over others in a particular span.}
\begin{tabular}{|l|l|l|p{7cm}|}	\hline
Director 		& Film & Year & Actors \\ \hline
Hrishikesh Mukherjee 	&	Abhimaan	&	1973	&	Amitabh Bachchan, Jaya Badhuri, Bindu		\\ \hline
Hrishikesh Mukherjee 	&	Chupke Chupke	&	1975	&	Dharmendra, Amitabh Bachchan, Jaya Badhuri, Sharmila Tagore		\\ \hline
Hrishikesh Mukherjee 	&	Mili	&	1975	&	Amitabh Bachchan, Jaya Badhuri, Ashok Kumar  	\\ \hline
Yash Chopra	&	Deewaar	&	1975	&	Amitabh Bachchan, Shashi Kapoor, Neetu Singh, Parveen Babi	\\ \hline
Yash Chopra	&	Kabhie Kabhie	&	1976	&	Amitabh Bachchan, Raakhee, Rishi Kapoor, Waheeda Rehman, Shashi Kapoor, Neetu Singh		\\ \hline
\end{tabular}
\end{center}
\label{director}
\end{supptable}
\section*{Data pertaining to events in the society}
The events in the society which have been related with decrease in randomness of the five datasets (Table 1 and Fig. 3 of manuscript) are enlisted in Table S2.
\begin{supptable}[h]
\begin{center}
\caption{Events in society which are related with decrease in $L_{0(w)}$.}
\begin{tabular}{|l|l|}	\hline
Year(s) & Event \\  \hline 
1962 & Sino-Indian War \\  \hline 
1971 & Indo-Pakistani War  \\  \hline
1984 & Operation Blue Star  \\  \hline
1984 & Operation Meghdoot \\  \hline
1987-90 & Sri Lankan Civil War \\  \hline
1988 &  Operation Cactus \\  \hline
1986-89 & Hindu-Muslim Communal Riots in India \cite{communal_riots} \\  \hline
1986 & Film industry went on strike due to contentious issues with state government \cite{No_Filmfare} \\  \hline 
\end{tabular}
\end{center}
\label{events}
\end{supptable}
\section*{Established actors of regional film industries}
Table S2 enlists the established actors of regional film industries who appear in relatively high $\beta_c$ yet high CC zone.
\begin{supptable}[h]
\begin{center}
\caption{List of established actors from regional film industries who have proved their realm in Bollywood.}
\begin{tabular}{|l|l|p{10cm}|}	\hline
Actor &	Span & Recognition \\  \hline
Kamal Haasan	& 1959-Present		& Indian film actor, screenwriter, director, producer, playback singer, choreographer and lyricist who works in the Tamil, Malayalam and films; has won four National Film Awards and 19 Filmfare Awards; received the Padma Shri (1990) and the Padma Bhushan (2014).     \\  \hline 
Padmini		& 1948-2002		& She acted in the Tamil, Telugu, Malayalam and Hindi language films; won Filmfare Award for Best Supporting Actress in 1966, Tamil Nadu state government awards and "Best Classical Dancer Award" from Moscow Youth Festival.   \\  \hline 
Manorama	& 1958-Present		& Appeared in more than 1000 films; won Padma Shri and regional film awards.     \\  \hline 
Uday Kumar	& 1956-1983		& Renowned as a Kannada film actor, producer and writer; won five National awards and State awards.    \\  \hline 
Jairaj		& 1929-1995		& Renowned Telugu film actor, producer and director; recipient of Dadasaheb Phalke award and Padma Bhushan.    \\  \hline 
\end{tabular}
\end{center}
\label{actors_regional}
\end{supptable}
\section*{Box Office and relevant statistics}
The Box Office India is a website {\it www.boxofficeindia.com} located in India and Texas, United States that tracks box office revenue in a systematic, algorithmic way of, especially, Bollywood films. It was launched in 2003. Its server is located in Houston, US. It also creates an overall week chart for domestic collections and update final worldwide gross of Hindi movies. It updates opening and final figures of overseas collection of Hindi films from various countries as well as the collection of Hollywood films in India.
The movie ``Mangal Pandey: The Rising" was declared below average at Box Office owing to its gross turnover of 550.1 million which is much less than the double of the movie budget of 380 million \cite{Mangal_pandey}.
\section*{Data pertaining to relation between change in degree order and success for Bollywood stars}
We first sort all the actors in descending order of their weighted degrees in all datasets. Then we pick top ten actors from each span who have the highest number of Filmfare award nominations to their credit. We find the ratio of their positions in degree order in unweighted networks to those of weighted networks for these 10 actors from each dataset denoted by $X$. The value of this ratio above 1 means that the actor has co-acted with his/her co-actors relatively more frequently than the rest of the actors of that span. This ratio $X$ is calculated for each actor in the dataset under consideration, the previous span and the succeeding span. In order to assess the impact of co-actors having acted more frequently on the success of the actors, we calculate the ratio of $X$ for each actor in consecutive time spans and also the ratio of their award nomanations (denoted by $A$) in respective spans (provided in Tables S3-S12). Occurrence of ratio of $X$ and ratio of $A$ in respective consecutive spans for any actor greater than $1$ simultaneously indicates that working in closed circles are implicative in success in Bollywood.
\begin{supptable}[h]
\begin{center}
\caption{List of lead actors in 03-07 dataset rising up in degree order as we proceed from unweighted Bollywood network analysis to weighted analysis. '-' indicates that the actor has not been nominated even once in that span. $X$ refers to the ratio of the position of each actor in degree order in unweighted network to that in weighted network. $A$ refers to the number of times an actor is nominated in a particular span.}
\begin{tabular}{|l|l|l|l|l|}	\hline
Actors 	 & \multicolumn{1}{|p{1.2 cm}|}{\centering $X_{03-07}$/ \\ $X_{98-02}$} & \multicolumn{1}{|p{1.2 cm}|}{\centering $X_{08-12}$/ \\ $X_{03-07}$} & \multicolumn{1}{|p{1.2 cm}|}{\centering $A_{03-07}$/ \\ $A_{98-02}$} & \multicolumn{1}{|p{1.2 cm}|}{\centering $A_{08-12}$/ \\ $A_{03-07}$}  \\ \hline
Shahrukh Khan	&	1.47	&	0.97	&	1.12	&	0.78	\\ \hline
Abhishek Bachchan &	0.73	&	1.67	&	-	&	0.43	\\ \hline
Ajay Devgn	&	1.61	&	0.92	&	2	&	0.33	\\ \hline
Hrithik Roshan	&	1.12	&	0.94	&	1.25	&	0.59	\\ \hline
Akshay Kumar	&	0.9	&	0.97	&	2	&	0.5	\\ \hline
Sanjay Dutt	&	1.2	&	1.26	&	2	&	-	\\ \hline
Saif Ali Khan	&	1.06	&	0.97	&	2	&	0.25	\\ \hline
Aamir Khan	&	1.2	&	1.02	&	0.75	&	1	\\ \hline
Salman Khan	&	0.92	&	1.16	&	0.76	&	0.67	\\ \hline
John Abraham	&	-	&	1.04	&	-	&	-	\\ \hline
\end{tabular}
\end{center}
\label{lead_03-07}
\end{supptable}
\begin{supptable}[h]
\begin{center}
\caption{List of lead actors in 98-02 dataset rising up in degree order as we proceed from unweighted Bollywood network analysis to weighted analysis. '-' indicates that the actor has not been nominated even once in that span. $X$ refers to the ratio of the position of each actor in degree order in unweighted network to that in weighted network. $A$ refers to the number of times an actor is nominated in a particular span.}
\begin{tabular}{|l|p {1 cm}|p {1 cm}|p {1 cm}|p {1 cm}|}	\hline
Actors  & \multicolumn{1}{|p{1.2 cm}|}{\centering $X_{98-02}$/ \\ $X_{93-97}$} & \multicolumn{1}{|p{1.2 cm}|}{\centering $X_{03-07}$/ \\ $X_{98-02}$} & \multicolumn{1}{|p{1.2 cm}|}{\centering $A_{98-02}$/ \\$A_{93-97}$} & \multicolumn{1}{|p{1.2 cm}|}{\centering $A_{03-07}$/\\$A_{98-02}$}   \\ \hline
Shahrukh Khan	&	1.09	&	1.47	&	1.15	&	1.12	\\ \hline
Govinda		&	1.04	&	0.67	&	1.75	&	0.14	\\ \hline
Anil Kapoor	&	1.04	&	1.03	&	1.67	&	0.2	\\ \hline
Amitabh Bachchan &	0.94	&	0.6	&	5	&	2.22	\\ \hline
Manoj Bajpai	&	0.9	&	1.15	&	-	&	0.4	\\ \hline
Hrithik Roshan	&	-	&	1.14	&	-	&	1.25	\\ \hline
Aamir Khan	&	0.77	&	1.2	&	0.8	&	0.75	\\ \hline
Salman Khan	&	1.03	&	0.92	&	2	&	0.75	\\ \hline
Ajay Devgn	&	0.71	&	1.61	&	3.03	&	2	\\ \hline
Akshaye Khanna	&	1	&	1.04	&	-	&	-	\\ \hline
Suniel Shetty	&	0.93	&	0.61	&	3.03	&	-	\\ \hline
\end{tabular}
\end{center}
\label{lead_98-02}
\end{supptable}
\begin{supptable}[h]
\begin{center}
\caption{List of lead actors in 93-97 dataset rising up in degree order as we proceed from unweighted Bollywood network analysis to weighted analysis. '-' indicates that the actor has not been nominated even once in that span. $X$ refers to the ratio of the position of each actor in degree order in unweighted network to that in weighted network. $A$ refers to the number of times an actor is nominated in a particular span.}
\begin{tabular}{|l|p {1 cm}|p {1 cm}|p {1 cm}|p {1 cm}|}	\hline
Actors 		& \multicolumn{1}{|p{1.2 cm}|}{\centering $X_{93-97}$/ \\$X_{88-92}$} & \multicolumn{1}{|p{1.2 cm}|}{\centering $X_{98-02}$/ \\ $X_{93-97}$} & \multicolumn{1}{|p{1.2 cm}|}{\centering $A_{93-97}$/\\ $A_{88-92}$} & \multicolumn{1}{|p{1.2 cm}|}{\centering $A_{98-02}$/ \\$A_{93-97}$} \\ \hline
Shahrukh Khan	&	0.64	&	1.09	&	-	&	1.15	\\ \hline
Nana Patekar	&	1.07	&	0.81	&	3.03	&	-	\\ \hline
Aamir Khan	&	0.98	&	0.77	&	1	&	0.8	\\ \hline
Jackie Shroff	&	0.74	&	0.85	&	5	&	0.4	\\ \hline
Govinda		&	1.39	&	1.04	&	-	&	1.75	\\ \hline
Anil Kapoor	&	0.88	&	1.04	&	1	&	1.67	\\ \hline
Salman Khan	&	1.15	&	1.03	&	2	&	2	\\ \hline
Sunny Deol	&	0.69 &	1.16	&	2	&	0.5	\\ \hline
Saif Ali Khan	&	-	&	0.96	&	-	&	1	\\ \hline
Ajay Devgn	&	1.14	&	0.71	&	1	&	3.03	\\ \hline
\end{tabular}
\end{center}
\label{lead_93-97}
\end{supptable}
\begin{supptable}[h]
\begin{center}
\caption{List of lead actors in 88-92 dataset rising up in degree order as we proceed from unweighted Bollywood network analysis to weighted analysis. '-' indicates that the actor has not been nominated even once in that span. $X$ refers to the ratio of the position of each actor in degree order in unweighted network to that in weighted network. $A$ refers to the number of times an actor is nominated in a particular span.}
\begin{tabular}{|l|l|l|l|l|}	\hline
Actors 		& \multicolumn{1}{|p{1.2 cm}|}{\centering $X_{88-92}$/ \\ $X_{83-87}$} & \multicolumn{1}{|p{1.2 cm}|}{\centering $X_{93-97}$/ \\$X_{88-92}$} & \multicolumn{1}{|p{1.2 cm}|}{\centering $A_{88-92}$/ \\$A_{83-87}$} & \multicolumn{1}{|p{1.2 cm}|}{\centering $A_{93-97}$/ \\$A_{88-92}$}    \\ \hline
Aamir Khan	&	1.05	&	0.98	&	-	&	1		\\ \hline
Amitabh Bachchan &	1.05	&	1.04	&	1	&	0.25		\\ \hline
Anil Kapoor	&	1.18	&	0.88	&	1.49	&	1		\\ \hline
Nana Patekar	&	0.98	&	1.07	&	-	&	3.03		\\ \hline
Mithun Chakraborty &	0.66	&	1.29	&	-	&	1		\\ \hline
Sunny Deol	&	1.33	&	0.69	&	1	&	2		\\ \hline
Sanjay Dutt	&	0.93	&	0.88	&	-	&	1		\\ \hline
Salman Khan	&	-	&	1.15	&	-	&	2		\\ \hline
Jackie Shroff	&	0.87	&	0.74	&	-	&	5		\\ \hline
Ajay Devgn	&	-	&	1.14	&	-	&	1		\\ \hline
\end{tabular}
\end{center}
\label{lead_88-92}
\end{supptable}
\begin{supptable}[h]
\begin{center}
\caption{List of lead actors in 83-87 dataset rising up in degree order as we proceed from unweighted Bollywood network analysis to weighted analysis. '-' indicates that the actor has not been nominated even once in that span. $X$ refers to the ratio of the position of each actor in degree order in unweighted network to that in weighted network. $A$ refers to the number of times an actor is nominated in a particular span.}
\begin{tabular}{|l|l|l|l|l|}	\hline
Actors 		& \multicolumn{1}{|p{1.2 cm}|}{\centering $X_{83-87}$/ \\$X_{78-82}$} & \multicolumn{1}{|p{1.2 cm}|}{\centering $X_{88-92}$/ \\$X_{83-87}$} & \multicolumn{1}{|p{1.2 cm}|}{\centering $A_{83-87}$/ \\$A_{78-82}$} & \multicolumn{1}{|p{1.2 cm}|}{\centering $A_{88-92}$/ \\$A_{83-87}$}    \\ \hline
Naseruddin Shah	&	0.89	&	0.89	&	1.67	&	-		\\ \hline
Amitabh Bachchan &	0.66	&	1.05	&	0.4	&	1		\\ \hline
Kamal Hassan	&	0.78	&	1.29	&	3.03	&	-		\\ \hline
Raj Babbar	&	1.09	&	1.32	&	2	&	-		\\ \hline
Rishi Kapoor	&	0.75	&	1.16	&	2	&	0.5		\\ \hline
Anil Kapoor	&	-	&	1.18	&	-	&	1.49		\\ \hline
Sanjeev Kumar	&	0.92	&	-	&	0.4	&	-		\\ \hline
Dilip Kumar	&	1.03	&	1.05	&	-	&	0.5		\\ \hline
Dharmendra	&	1.09	&	1.59	&	-	&	-		\\ \hline
Shashi Kapoor	&	1.52	&	1	&	1	&	-		\\ \hline
\end{tabular}
\end{center}
\label{lead_83-87}
\end{supptable}
\begin{supptable}[h]
\begin{center}
\caption{List of lead actors in 78-82 dataset rising up in degree order as we proceed from unweighted Bollywood network analysis to weighted analysis. '-' indicates that the actor has not been nominated even once in that span. $X$ refers to the ratio of the position of each actor in degree order in unweighted network to that in weighted network. $A$ refers to the number of times an actor is nominated in a particular span.}
\begin{tabular}{|l|l|l|l|l|l|l|}	\hline
Actors 		& \multicolumn{1}{|p{1.2 cm}|}{\centering $X_{78-82}$/ \\$X_{73-77}$} & \multicolumn{1}{|p{1.2 cm}|}{\centering $X_{83-87}$/ \\$X_{78-82}$} & \multicolumn{1}{|p{1.2 cm}|}{\centering $A_{78-82}$/ \\$A_{73-77}$} & \multicolumn{1}{|p{1.2 cm}|}{\centering $A_{83-87}$/ \\ $A_{78-82}$} \\ \hline
Amitabh Bachchan &	1.54	&	0.66	&	2.5	&	0.4		\\ \hline
Sanjeev Kumar 	&	0.86	&	0.92	&	1	&	0.4		\\ \hline
Rajesh Khanna	&	0.94	&	2.04	&	0.5	&	0.33		\\ \hline
Vinod Khanna	&	1.15	&	0.63	&	1.49	&	-		\\ \hline
Naseruddin Shah	&	-	&	0.89	&	-	&	1.67		\\ \hline
Shatrughan Sinha &	1.39	&	2.08	&	2	&	-		\\ \hline
Vinod Mehra	&	1.41	&	0.65	&	-	&	0.5		\\ \hline
Raj Babbar	&	-	&	1.11	&	-	&	2		\\ \hline
Kamal Hassan	&	1.06	&	0.78	&	-	&	3.03		\\ \hline
Rishi Kapoor	&	1.15	&	0.75	&	1	&	2		\\ \hline
\end{tabular}
\end{center}
\label{lead_78-82}
\end{supptable}
\begin{supptable}[h]
\begin{center}
\caption{List of lead actors in 73-77 dataset rising up in degree order as we proceed from unweighted Bollywood network analysis to weighted analysis. '-' indicates that the actor has not been nominated even once in that span. $X$ refers to the ratio of the position of each actor in degree order in unweighted network to that in weighted network. $A$ refers to the number of times an actor is nominated in a particular span.}
\begin{tabular}{|l|l|l|l|l|l|l|}	\hline
Actors 		& \multicolumn{1}{|p{1.2 cm}|}{\centering $X_{73-77}$/ \\ $X_{68-72}$} & \multicolumn{1}{|p{1.2 cm}|}{\centering $X_{78-82}$/ \\$X_{73-77}$} & \multicolumn{1}{|p{1.2 cm}|}{\centering $A_{73-77}$/ \\$A_{68-72}$} & \multicolumn{1}{|p{1.2 cm}|}{\centering $A_{78-82}$/ \\$A_{73-77}$ } \\ \hline
Rajesh Khanna	&	1.03	&	0.94	&	1.2	&	0.5				\\ \hline
Sanjeev Kumar 	&	1.61	&	0.86 &	2.5	&	1				\\ \hline
Amitabh Bachchan &	0.95	&	1.54	&	4	&	2.5				\\ \hline
Manoj Kumar	&	1.02	&	-	&	1.49	&	-				\\ \hline
Dilip Kumar	&	1.07	&	0.97	&	0.5	&	-				\\ \hline
Shashi Kapoor	&	1.06	&	0.5	&	-	&	-				\\ \hline
Dharmendra	&	0.67	&	1.75	&	2	&	-				\\ \hline
Vinod Khanna	&	2.08	&	0.57	&	-	&	1.49				\\ \hline
Rishi Kapoor	&	-	&	1.15	&	-	&	1				\\ \hline
Shatrughan Sinha &	0.74	&	1.61	&	1	&	2				\\ \hline
\end{tabular}
\end{center}
\label{lead_73-77}
\end{supptable}
\begin{supptable}[h]
\begin{center}
\caption{List of lead actors in 68-72 dataset rising up in degree order as we proceed from unweighted Bollywood network analysis to weighted analysis. '-' indicates that the actor has not been nominated even once in that span. $X$ refers to the ratio of the position of each actor in degree order in unweighted network to that in weighted network. $A$ refers to the number of times an actor is nominated in a particular span.}
\begin{tabular}{|l|l|l|l|l|l|l|}	\hline
Actors 		& \multicolumn{1}{|p{1.2 cm}|}{\centering $X_{68-72}$/ \\ $X_{63-67}$} & \multicolumn{1}{|p{1.2 cm}|}{\centering $X_{73-77}$/ \\$X_{68-72}$} & \multicolumn{1}{|p{1.2 cm}|}{\centering $A_{68-72}$/ \\$A_{63-67}$} & \multicolumn{1}{|p{1.2 cm}|}{\centering $A_{73-77}$/ \\$A_{68-72}$} \\ \hline
Rajesh Khanna 	&	1.15	&	1.03	&	-	&	1.2		\\ \hline
Dilip Kumar 	&	0.97	&	1.07	&	2	&	0.5		\\ \hline
Sanjeev Kumar	&	0.71	&	1.61	&	-	&	2.5    	\\ \hline
Manoj Kumar	&	0.68	&	1.02	&	-	&	1.49		\\ \hline
Feroz Khan	&	1.03	&	1.03	&	-	&	0.5		\\ \hline
Dharmendra 	&	1.4	&	0.67	&	0.5	&	2		\\ \hline
Shammi Kapoor	&	0.97	&	0.92	&	1	&	-		\\ \hline
Sunil Dutt	&	1.07	&	0.93	&	1	&	-		\\ \hline
Amitabh Bachchan &	-	&	0.95	&	-	&	4		\\ \hline
Shatrughan Sinha &	-	&	0.74	&	-	&	1		\\ \hline
\end{tabular}
\end{center}
\label{lead_68-72}
\end{supptable}
\begin{supptable}[h]
\begin{center}
\caption{List of lead actors in 63-67 dataset rising up in degree order as we proceed from unweighted Bollywood network analysis to weighted analysis. '-' indicates that the actor has not been nominated even once in that span. $X$ refers to the ratio of the position of each actor in degree order in unweighted network to that in weighted network. $A$ refers to the number of times an actor is nominated in a particular span.}
\begin{tabular}{|l|l|l|l|l|l|l|}	\hline
Actors 		& \multicolumn{1}{|p{1.2 cm}|}{\centering $X_{63-67}$/ \\ $X_{58-62}$} & \multicolumn{1}{|p{1.2 cm}|}{\centering $X_{68-72}$/ \\$X_{63-67}$} & \multicolumn{1}{|p{1.2 cm}|}{\centering $A_{63-67}$/ \\$A_{58-62}$} & \multicolumn{1}{|p{1.2 cm}|}{\centering $A_{68-72}$/ \\$A_{63-67}$} \\ \hline
Raj Kapoor 	&	1.15	&	0.94	&	0.25	&	-		\\ \hline
Ashok Kumar 	&	0.84	&	0.57	&	-	&	1		\\ \hline
Raaj Kumar	&	1.09	&	0.97	&	-	&	-	    	\\ \hline
Mehmood		&	1.03	&	0.71	&	6	&	1.33		\\ \hline
Dilip Kumar	&	0.91	&	0.97	&	0.5	&	2		\\ \hline
Sunil Dutt		&	1	&	1.07	&	-	&	0.5		\\ \hline
Dharmendra	&	0.81	&	1.41	&	-	&	0.5		\\ \hline
Dev Anand	&	1.15	&	1.14	&	0.33	&	-		\\ \hline
Guru Dutt		&	1.03	&	-	&	-	&	-		\\ \hline
Shammi Kapoor &	1.08	&	0.97	&	-	&	1		\\ \hline
\end{tabular}
\end{center}
\label{lead_63-67}
\end{supptable}
\begin{supptable}[h]
\begin{center}
\caption{List of lead actors in 58-62 dataset rising up in degree order as we proceed from unweighted Bollywood network analysis to weighted analysis. '-' indicates that the actor has not been nominated even once in that span. $X$ refers to the ratio of the position of each actor in degree order in unweighted network to that in weighted network. $A$ refers to the number of times an actor is nominated in a particular span.}
\begin{tabular}{|l|l|l|l|l|l|l|}	\hline
Actors 		& \multicolumn{1}{|p{1.2 cm}|}{\centering $X_{58-62}$/ \\ $X_{53-57}$} & \multicolumn{1}{|p{1.2 cm}|}{\centering $X_{63-67}$/ \\$X_{58-62}$} & \multicolumn{1}{|p{1.2 cm}|}{\centering $A_{58-62}$/ \\$A_{53-57}$} & \multicolumn{1}{|p{1.2 cm}|}{\centering $A_{63-67}$/ \\$A_{58-62}$} \\ \hline
Raj Kapoor 	&	0.84	&	1.15	&	4	&	0.25		\\ \hline
Dilip Kumar 	&	1.58	&	0.91	&	1.33	&	0.5		\\ \hline
Dev Anand	&	0.79	&	1.15	&	-	&	0.33 	\\ \hline
Sohrab Modi	&	0.98	&	0.97	&	-	&	-		\\ \hline
Mehmood		&	1.79	&	1.03	&	-	&	6		\\ \hline
\end{tabular}
\end{center}
\label{lead_58-62}
\end{supptable}

\section*{New comers and their co-actors}
Table S13 enlists the high degree co-actors of the new comers of a particular span.
\begin{supptable}
\begin{center}
\caption{High degree co-actors of the new comers of the industry in 08-12 dataset.}
\begin{tabular}{|c|c|c|}    \hline
New comer	& High degree co-actor	& Degree order of co-actor	\\ \hline
Xia Yu		& 	Anupam Kher	&	1		\\ \hline
Suman Negi	&      Viju Khote	&	49		\\ \hline
Ahad Khan	&      Tinu Anand	&	29		\\ \hline
Jitu Savlani	&      Rati Agnihotri	&	76		\\ \hline
Anil Kumble	&      Anupam Kher	&	1		\\ \hline
Charu Sharma	&      Anupam Kher	&	1		\\ \hline
Himayat Ali	&      Govind Namdev	&	8		\\ \hline
Amandeep Singh Bakshi	&      Paresh Rawal	&	16		\\ \hline
Jacqueline Grewal	&     Boman Irani 	&	27		\\ \hline
Chinmay Patwardhan	&      Boman Irani	&	27		\\ \hline
Amrit Maghera	&      Boman Irani	&	27		\\ \hline
Kamal Rashid Khan	&      Avtar Gill	&	30		\\ \hline
Vinita Malik	&      Dinesh Hingoo	&	35		\\ \hline
Sweety Chhabra	&      Dinesh Hingoo	&	35		\\ \hline
Sachin Khurana	&      Ranvir Shorey	&	39		\\ \hline
''	&     Vinay Pathak 	&	45		\\ \hline
''	&      Saurabh Shukla	&	24		\\ \hline
Ritu Vasishtha	&      Mushtaq Khan	&	7		\\ \hline
Amrit Kaur Chawla	&     Mushtaq Khan 	&	7		\\ \hline
\end{tabular}
\begin{flushleft} New comers are the actors who have acted in just one movie in a particular span. The second column consists of the names of the high degree co-actors these new comers have co-acted with. The third column consists of the positions of the high degree co-actors in degree sequence (Actors having their position within top 50 in the dataset consisting of 3611 nodes have been considered to have high degree.)
\end{flushleft}
\end{center}
\end{supptable}

\end{document}